\documentclass[runningheads]{llncs}

\usepackage{amssymb}
\usepackage{multicol}
\usepackage{multirow}
\usepackage[T1]{fontenc}
\usepackage{graphicx}
\begin{document}
\title{Two-Stage Approach for Brain MR Image Synthesis: 2D Image Synthesis and 3D Refinement}
\titlerunning{Two-Stage Approach for Brain MR Image Synthesis}
%
\author{Jihoon Cho\thanks{J. Cho and S. Park contributed equally to this work as first authors.}\orcidID{0000-0002-8619-9481} \and
Seunghyuck Park*\orcidID{0009-0005-9508-6664} \\ \and
Jinah Park\orcidID{0000-0003-4676-9862}}
\authorrunning{J. Cho et al.}
%
\institute{School of Computing, Korea Advanced Institute of Science and Technology, \\ Daejeon 34141, Republic of Korea \\
\email{\{zinic,simonp6,jinahpark\}@kaist.ac.kr}}
\maketitle              
\begin{abstract}
Despite significant advancements in automatic brain tumor segmentation methods, their performance is not guaranteed when certain MR sequences are missing. Addressing this issue, it is crucial to synthesize the missing MR images that reflect the unique characteristics of the absent modality with precise tumor representation. Typically, MRI synthesis methods generate partial images rather than full-sized volumes due to computational constraints. This limitation can lead to a lack of comprehensive 3D volumetric information and result in image artifacts during the merging process. In this paper, we propose a two-stage approach that first synthesizes MR images from 2D slices using a novel intensity encoding method and then refines the synthesized MRI. The proposed intensity encoding reduces artifacts when synthesizing MRI on a 2D slice basis. Then, the \textit{Refiner}, which leverages complete 3D volume information, further improves the quality of the synthesized images and enhances their applicability to segmentation methods. Experimental results demonstrate that the intensity encoding effectively minimizes artifacts in the synthesized MRI and improves perceptual quality. Furthermore, using the \textit{Refiner} on synthesized MRI significantly improves brain tumor segmentation results, highlighting the potential of our approach in practical applications.

\keywords{Image Synthesis  \and Intensity Encoding \and Refiner}
\end{abstract}
\section{Introduction}
A brain tumor is one of the deadliest types of cancer, and accurate brain tumor segmentation is a crucial step in diagnosis and treatment planning. However, the segmentation process of the tumor and its sub-regions within brain tissue is highly labor intensive and time-consuming because of the varying visible characteristics in different magnetic resonance imaging (MRI). Recently, automatic segmentation methods have emerged as a promising solution, showing outstanding performance with rapid processing time~\cite{cho2022hybrid,isensee2021nnu,FeTS}. They have been designed to handle four MR sequences: T1-weighted MRI (T1), T1-weighted MRI with contrast enhancement (T1ce), T2-weighted MRI (T2), and T2-weighted Fluid-Attenuated Inversion Recovery MRI (FLAIR). These MR sequences contribute to the achievement of optimal performance by providing unique features of each modality.
However, in practical scenarios, acquiring all four modalities is difficult due to several practical issues such as time constraints, image artifacts, and limited access to imaging equipment. This presents a significant challenge in the application of automatic segmentation methods in clinical practice. 

A common approach to impute missing data is to synthesize missing MRI. However, synthesizing high-resolution 3D images at once is impractical in most hardware systems due to its high complexity and substantial computational cost. Consequently, slice-based image synthesis methods~\cite{cho2024unified,dalmaz2022resvit,liu2023one} have emerged as a viable alternative, yet suffer two significant drawbacks due to their 2D slice-based nature. First, when stacking 2D slices to reconstruct a 3D volume, intensity inconsistencies between slices can lead to stripe artifacts when viewed in slices of a different axis. Second, since the image synthesis relies only on intra-slice information without sufficient 3D structural information, tumor segmentation results with the synthesized images show suboptimal performance.

To address these problems, we employ the two-stage approach that benefits from both 2D and 3D methods. In the first stage, we synthesize axial slices of a missing MRI and then stack these slices to reconstruct a 3D volume. To minimize the intensity inconsistencies between the slices, we propose a novel intensity encoding method. Integrating intensity encoding with modality encoding effectively controls the intensity level regardless of the target modality.
In the second stage, we propose the \textit{Refiner} which improves the output from the first stage by incorporating comprehensive 3D information. The \textit{Refiner}, with its 3D receptive field, improves the tumor representation of synthesized MRI by leveraging 3D information of the tumor region obtained from available MR sequences. We conducted experiments with the ASNR-MICCAI BraTS MRI Synthesis Challenge (BraSyn)~\cite{li2023braintumorsegmentationbrats}. The experimental results show the effectiveness of the proposed intensity encoding method in enhancing the perceptual quality of synthesized MRI, as well as the capability of the proposed \textit{Refiner} in significantly improving the representation of tumor region and our approach achieved first place in the 2024 BraSyn challenge.

\section{Method}

Our proposed two-stage approach consists of 2D-based MR image synthesis and 3D-based MR image refinement. We focus on a detailed image synthesis with intensity consistency in the first stage and a refinement of the synthesized MRI with tumor representation in the second stage. Detailed descriptions of each stage are provided in the following subsections.

\subsection{First Stage: 2D-based MR Image Synthesis}
We use the state-of-the-art MR image synthesis method \textit{HF-GAN}~\cite{cho2024unified} as a baseline to synthesize 2D MR images. \textit{HF-GAN} consists of the hybrid fusion encoder, the channel attention-based feature fusion module, the modality infuser, and a CNN decoder. First, the hybrid fusion encoder accepts the acquired MR sequences as inputs. The extracted feature representations are projected into a common latent space through a channel attention-based feature fusion module. Lastly, the modality infuser transforms feature representations in a common latent space into a target latent space with modality encoding, subsequently generating the missing target modality via a CNN decoder. 

\begin{figure}[t]
\begin{center}
\includegraphics[width=1.0\linewidth]{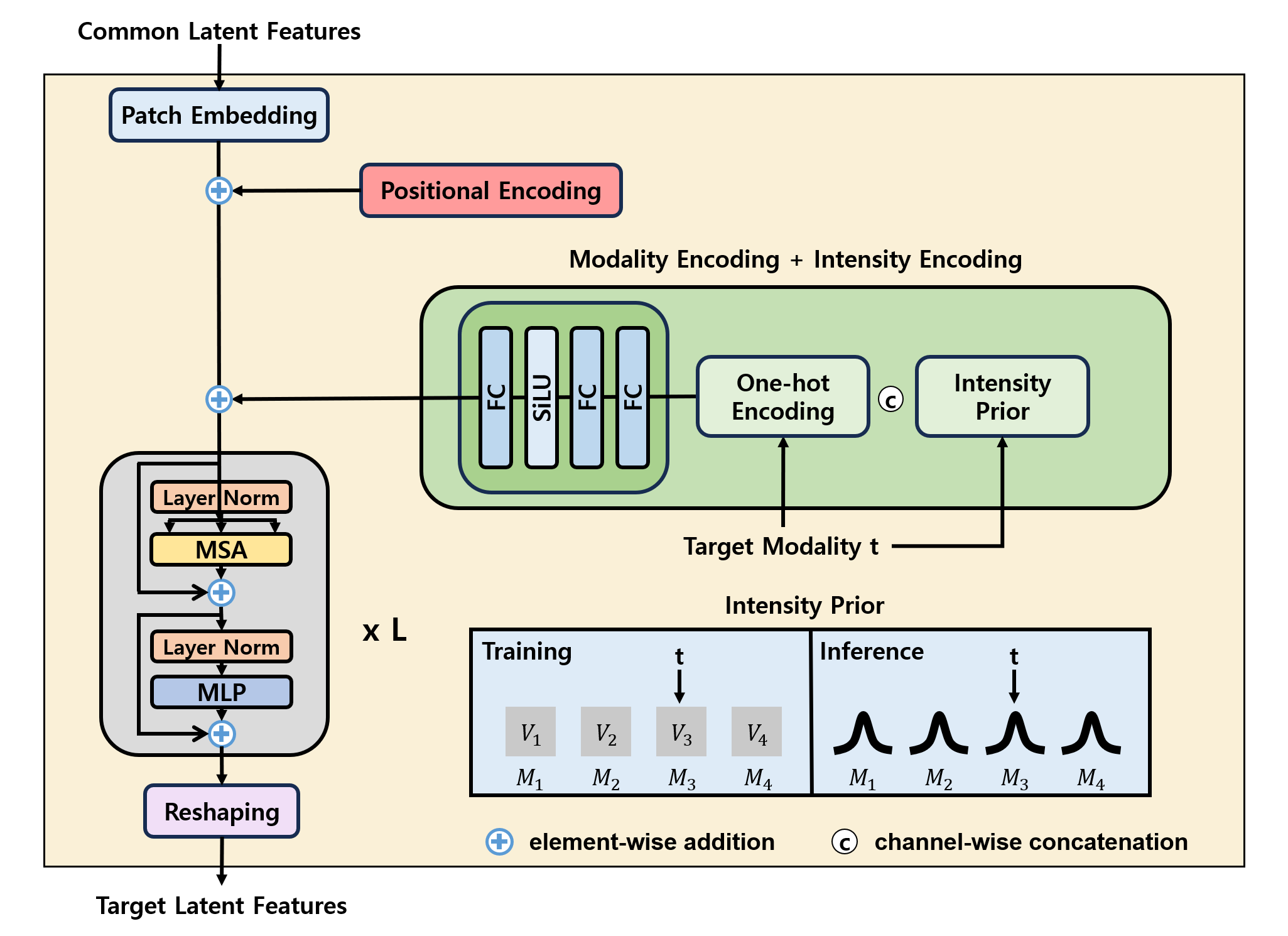}
\end{center} 
\caption{The structure of modality infuser with intensity encoding. The target intensity and target modality are jointly encoded. In the training phase, the target intensity is directly computed from the target MR sequence. In the inference phase, the intensity prior is pre-computed from the training dataset and the target intensity is sampled from this distribution.}
\label{fig:ie}
\end{figure} 

\subsubsection{Intensity Encoding}
A 3D volume of the missing MR sequence can be reconstructed by slice-by-slice synthesis. However, this naive approach suffers from a lack of interaction between the slices, leading to intensity inconsistency caused by the high variability in MRI intensity. To minimize inter-slice inconsistency, we propose a novel intensity encoding method that can be integrated into the modality infuser as shown in Figure~\ref{fig:ie}. In order to maintain a consistent intensity level, the target intensity is conditioned through the intensity encoding along with the modality encoding. To accurately reflect the intensity level of the volume irrespective of the slice's position or tumor presence, we select the median intensity of brain area in MRI, generally corresponding to the intensity between gray matter and white matter. For the inference phase where the MRI is missing and the median intensity cannot be obtained, we precompute the normal distribution of the median intensity using the training dataset and employ this distribution for statistical sampling.

\subsection{Second Stage: 3D-based MRI Refinement}
\begin{figure}[t]
\begin{center}
\includegraphics[width=1.0\linewidth]{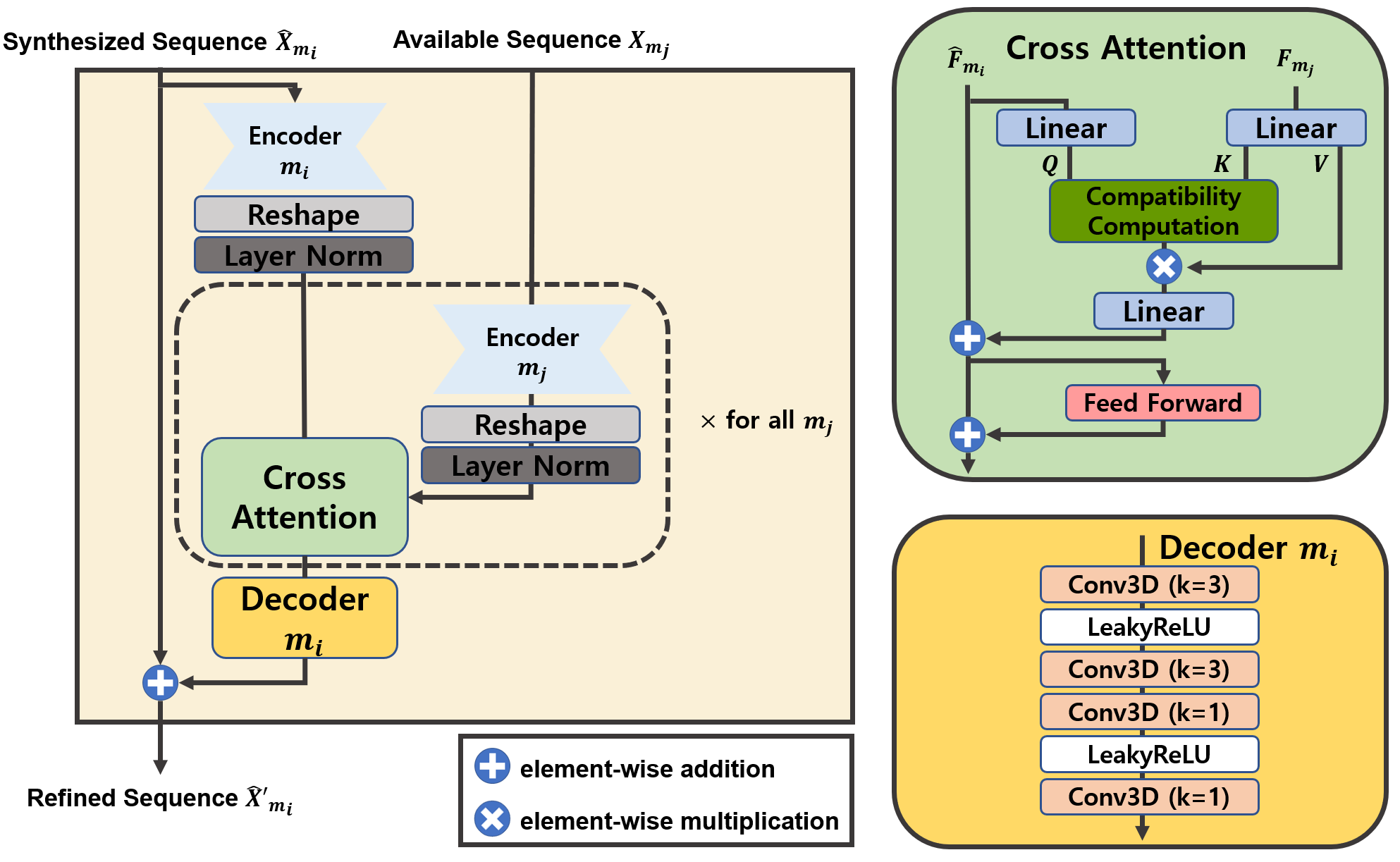}
\end{center} 
\caption{The structure of \textit{Refiner}. The feature representation of synthesized MR image, $\hat{F}_{m_i}$, and of all available MR image, $F_{m_j}$, are obtained using the corresponding encoder. Then, voxel-wise cross-attention is performed between $\hat{F}_{m_i}$ and each of $F_{m_j}$ sequentially. Finally, the refined feature representation is decoded into the image space and added to the input MR image $\hat{X}_{m_i}$ to produce the final output $\hat{X}_{m_i}'$.}
\label{fig:enc}
\end{figure} 

The \textit{Refiner}'s objective is to enhance the quality of the MR image synthesized on a 2D-slice basis by improving tumor representation by incorporating 3D information from available MR images.

The \textit{Refiner} is composed of three main components: the encoder, the element-wise cross-attention module, and the decoder as shown in Figure \ref{fig:enc}. The encoder is responsible for extracting both global and local features from the MR image. It encodes the synthesized MR image and three available images individually, resulting in a total of four feature representations. The element-wise cross-attention module then performs cross-attention in an element-wise manner on the features obtained from the encoder, refining the feature representation of the synthesized MR image. Lastly, the decoder reconstructs the MR image from the refined features.

\subsubsection{Encoder}

There are four modality-specific encoders that each have a simple U-net~\cite{ronneberger2015unetconvolutionalnetworksbiomedical} architecture. 
Given a synthesized MR sequence $\hat{X}_{m_i} \in \mathbb{R}^{1\times D\times H\times W} $, and available MR sequences $X_{m_j} \in \mathbb{R}^{1\times D\times H\times W} $, where $m_j \in \{T1, T2, FLAIR, T1ce\}$, the feature representations $\hat{F}_{m_i}$ and $ F_{m_j} \in \mathbb{R}^{C\times D\times H\times W}$ are obtained by passing each individual image to the corresponding encoder $Enc_i$, formulated as:

\begin{equation}
    \hat{F}_{m_i} = Enc_{m_i} (\hat{X}_{m_i}),
\end{equation}
\begin{equation}
    F_{m_j} = Enc_{m_j} (X_{m_j}).
\end{equation}

\subsubsection{Element-wise Cross-Attention Module}

The element-wise cross-attention module employs a mechanism similar to, yet distinct from, the conventional attention mechanisms. The main difference is that it performs cross-attention between voxels at the same spatial location in an element-wise manner.

The feature representation from the encoder in shape $[C\times D\times H\times W]$ is first reshaped into $[M\times C]$, where $M=D\times H\times W$, and the layer norm is applied.
Then, the query vector is extracted from $\hat{F}_{m_i}$, and the key and value vectors are extracted from $ F_{m_j}$ by linear projection. 

The extracted query and key vectors are used to calculate the compatibility between corresponding elements to serve as the weight vector. Since each element is in the form of a scalar instead of a vector, the dot product cannot be utilized to compute compatibility as in the conventional attention mechanism. Instead, we compute the element-wise absolute difference and apply a negative exponential such that the similarity is maximized (up to 1) when the values are identical and approaches 0 as the values diverge.

The resulting weight vector is then used in an element-wise product with the value vector and added to $\hat{F}_{m_i}$ to obtain the refined feature $\hat{F}_{m_i}'$. Lastly, a feedforward layer composed of 1x1 convolutions is applied to $\hat{F}_{m_i}'$ and added element-wise again to obtain further refined $\hat{F}_{m_i}''$.


This \textit{Refiner}'s process can be formulated as follows:

\begin{equation}
    Attention_{m_j} = (e^{ -|Q_{m_i} - K_{m_j}|}) \odot V_{m_j},
\end{equation}
\begin{equation}
    \hat{F}_{m_i}' = \hat{F}_{m_i} + Attention_{m_j}, 
\end{equation}
\begin{equation}
    \hat{F}_{m_i}'' = \hat{F}_{m_i}' + FF(\hat{F}_{m_i}').
\end{equation}

The element-wise cross-attention is performed on $\hat{F}_{m_i}$ pairing with all available $\hat{F}_{m_j}$ in a sequential manner with prefixed order.

\subsubsection{Decoder}

In the element-wise cross-attention module, each voxel is refined independently, which might lead to inconsistencies between adjacent voxels. To address this issue, the decoder first processes the features through several convolution layers with a kernel size of 3, ensuring that each voxel is consistent with its neighboring voxels. Lastly, a final 1x1 convolution is applied to decode the feature representation from the shape $[C\times D\times H\times W]$ to the shape $[1\times D\times H\times W]$. The output is then added to the input synthesized image as a refinement.

\section{Results}
\subsection{Dataset}

We experimented with the BraSyn 2024 dataset~\cite{asnr-dataset,Bakas2017,li2023braintumorsegmentationbrats,BRATS} consisting of 1,251 training subjects, 219 validation subjects, and 629 test subjects. Each training subject includes four MR sequences (T1, T2, FLAIR, T1ce) with tumor segmentation masks. The validation set also includes the four MRI modalities but does not provide tumor segmentation masks. All images have a size of 240x240x155 and have undergone skull-stripping and registration.

For the first stage, the data was normalized to the intensity range of [-1, 1] using linear scaling. 240x240-sized axial slices were used as input, but those slices with fewer than 2,000 brain pixels were excluded from the training set to ensure the quality of the training data.

For the second stage, each MRI image was processed with modality-specific z normalization, where only non-zero values were taken into account. 128x128x128 patches were then randomly cropped from the pre-processed MRI volumes as training input.

\subsection{Evaluation Metrics}
The perceptual quality of the synthesized MRI is evaluated using the structural similarity index measure (SSIM). For evaluation, each MRI is synthesized using the remaining three modalities, excluding the one that is synthesized. The SSIM score is then calculated by comparing the synthesized MRI with the ground truth MRI. 

For evaluating the quality of representation of the tumor region, the Dice score is used. Since the validation set does not include tumor segmentation masks, we predict segmentation labels of each validation subject using the FeTS model~\cite{FeTS} to use as pseudo-labels. Next, we used the FeTS method to perform brain tumor segmentation using the synthesized MR images as a part of the input. We then calculated the Dice score for three tumor sub-regions (whole tumor, enhancing tumor, tumor core) by comparing the segmentation results from the synthesized images with the pseudo-labels. We reported the 95\% Hausdorff distance (HD95) only for the hidden test set which has ground-truth labels.

\subsection{Experimental Results}

\begin{figure}[!t]
\begin{center}
\includegraphics[width=1.0\linewidth]{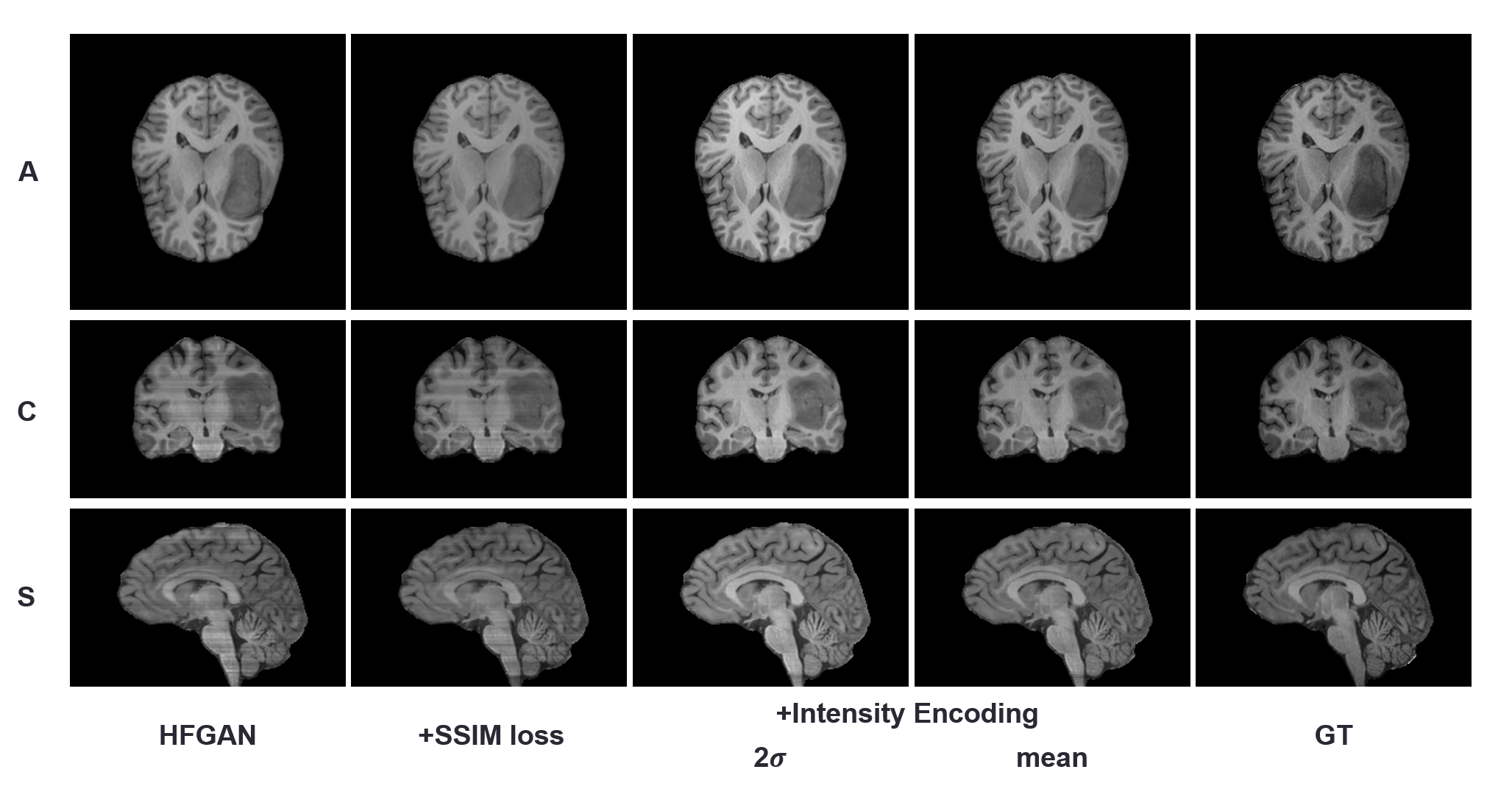}
\end{center} 
\caption{An example of 2D-based MR image synthesized results when T1 MRI is missing. A, C, and S represent axial, coronal and sagittal slices, respectively.}
\label{fig:first_eval}
\end{figure} 

\begin{table}[b]
\caption{Quantitative evaluation results of 2D-based missing MR image synthesis on the validation dataset. We evaluate synthesized 3D volumes, constructed from synthesized 2D slices, in comparison to actual MRIs. For intensity encoding, the \textit{mean} and \textit{2$\sigma$} indicate sampling approach using only the mean and clipping at 2$\sigma$, respectively. \textit{GT} is the upper bound of intensity encoding using the target intensity of ground truth.}\label{table:first_ssim}
\centering
\begin{tabular}{l|c|c|c|c|c}
\hline
\multirow{2}{*}{Method}  & \multicolumn{5}{c}{SSIM-Whole} \\ \cline{2-6}
                   & T1 & T2 & FLAIR & T1ce & Avg. \\ \hline \hline
HF-GAN~\cite{cho2024unified} & 0.9429 & 0.9421 & 0.9224 & 0.9230 & 0.9326 \\ 
+SSIM loss & 0.9477 & 0.9474 & 0.9301 & 0.9300 & 0.9388 \\ 
+Intensity Encoding (\textit{mean}) & \textbf{0.9553} & \textbf{0.9494} & \textbf{0.9353} & \textbf{0.9344} & \textbf{0.9436} \\ 
+Intensity Encoding (\textit{2$\sigma$}) & 0.9547 & 0.9460 & 0.9332 & 0.9331 & 0.9418 \\ \hline
+Intensity Encoding (\textit{GT}) & \textbf{0.9557} & \textbf{0.9538} & \textbf{0.9371} & \textbf{0.9372} & \textbf{0.9459} \\ \hline
\end{tabular}
\end{table}

\subsubsection{2D-based Missing MR Image Synthesis} 
Table~\ref{table:first_ssim} and Figure~\ref{fig:first_eval} present the quantitative and qualitative results on the synthesis of missing MR sequences. As discussed previously, stacking the synthesized 2D slices from the 2D-based image synthesis method has the intensity inconsistent issue while anatomical structures are consistent. As you can see in the first and second columns of Figure~\ref{fig:first_eval}, the stripe artifact occurred for the naive 2D-based approach even if SSIM loss is used. In contrast, by applying the proposed intensity encoding (as shown in the third and fourth columns), the stripe artifact was eliminated in the synthesized images. However, choosing a suitable sampling method for the target intensity becomes a new issue for intensity encoding. This is because while statistical sampling from the intensity prior can produce various intensity levels, it does not ensure exact intensity. The quantitative evaluation results of Table~\ref{table:first_ssim} indicate that employing deterministic sampling with the mean value for each intensity prior yields the highest SSIM scores across all missing MR cases. Specifically, it achieved an average SSIM score of 0.9436, representing an improvement of 0.0110 over HF-GAN and a minor degradation of only 0.0023 compared to the ground truth of the target intensity. Consequently, we choose the target intensity sampling method that utilizes only the mean value.

\subsubsection{3D-based Missing MRI Refinement}
\begin{figure}[!t]
\begin{center}
\includegraphics[width=1.0\linewidth]{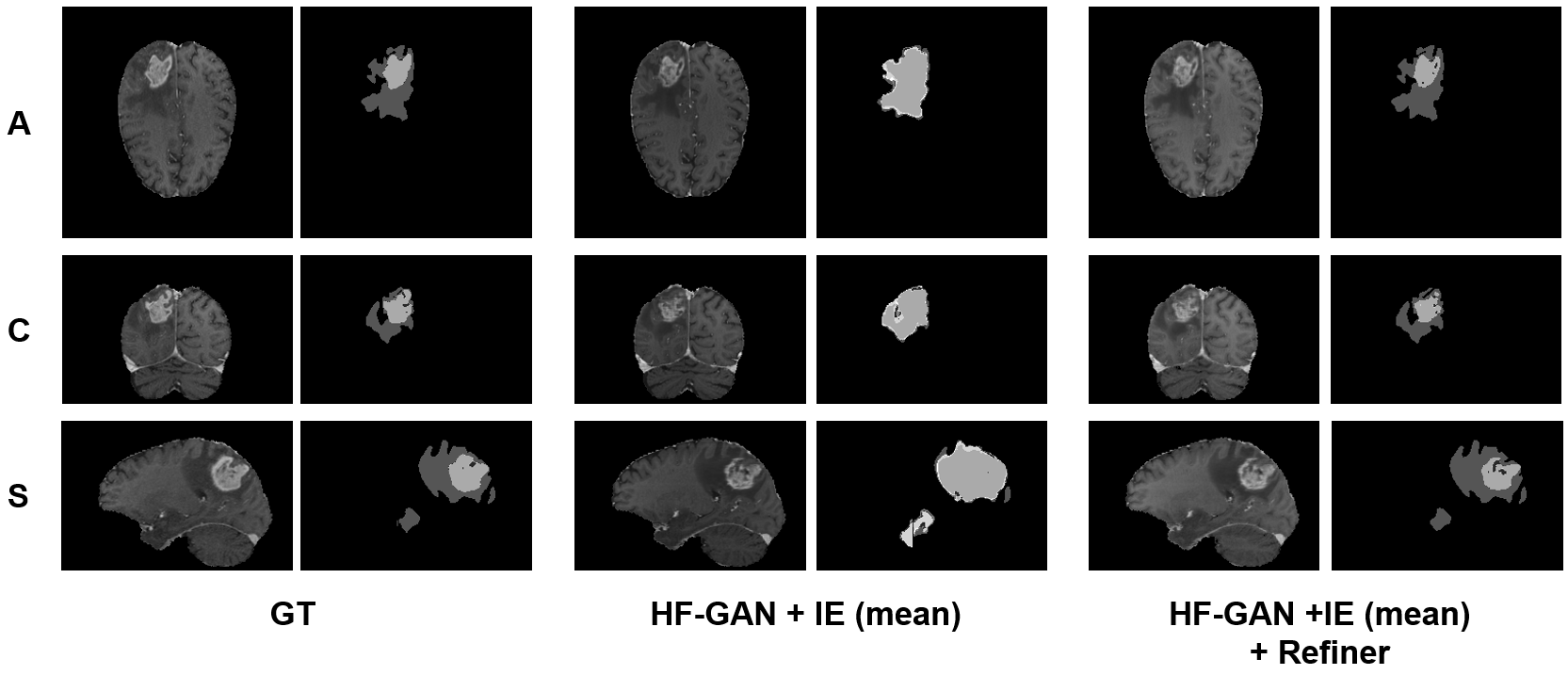}
\end{center} 
\caption{An example of synthesized output from the HF-GAN with Intensity Encoding (mean) and the refined output from the \textit{Refiner}. A, C, and S stand for axial, coronal, and sagittal slices, respectively, and next to each MR image is a segmentation prediction mask obtained using FeTS model.}
\label{fig:refiner_result}
\end{figure} 

\begin{table}[!b]
\caption{SSIM score between synthesized MR images and ground truth MR images on validation dataset before and after applying \textit{Refiner}. Using pseudo-label produced using FeTS model, SSIM scores for tumor region and healthy region are masked and calculated separately. IE (\textit{mean}) represents applying the intensity encoding method with sampling the target intensity as mean value.}
\label{tab1}
\centering
\begin{tabular}{l|c|c|c|c|c}
\hline
\multirow{2}{*}{Method} & \multicolumn{5}{c}{SSIM-Tumor region}  \\ \cline{2-6}
                    & T1 & T2 & FLAIR & T1ce & Avg \\ \hline \hline
HF-GAN + IE (\textit{mean}) &\textbf{0.9975} &\textbf{0.9973}  & 0.9975 & \textbf{0.9960} & \textbf{0.9970}  \\ 
+\textit{Refiner} & \textbf{0.9975} & 0.9971 & \textbf{0.9976} & 0.9959 & \textbf{0.9970}  \\ \hline \hline
\multirow{2}{*}{Method}  & \multicolumn{5}{c}{SSIM-Healthy region}  \\ \cline{2-6}
& T1 & T2 & FLAIR & T1ce & Avg \\ \hline \hline
HF-GAN + IE (\textit{mean}) & \textbf{0.9614} & \textbf{0.9533} & \textbf{0.9402} & \textbf{0.9425} & \textbf{0.9493} \\ 
+\textit{Refiner} & 0.9551 & 0.9449 & 0.9329 & 0.9372  & 0.9425\\ \hline
\end{tabular}
\label{tab:ssim}
\end{table}

The qualitative results are illustrated in Figure~\ref{fig:refiner_result}, and the quantitative results are illustrated in Table~\ref{tab:ssim} and Table~\ref{tab:Dice}. The results indicate that the \textit{Refiner} does not improve the perceptual quality of the synthesized MR images. 
As can be seen from Figure~\ref{fig:refiner_result}, the refined output does not show significant visual differences compared to the MR images synthesized from the first stage, yet little differences in intensity lead to a slightly lower SSIM score. Table ~\ref{tab:ssim} shows that the overall SSIM score on tumor region did not change, and the SSIM score on healthy region decreased by an average of 0.0068, with the most significant drop observed in the T2 modality, which saw a reduction of 0.0084. 
On the other hand, Table \ref{tab:Dice} shows that the \textit{Refiner} significantly improves tumor representation, increasing the dice score by more than 0.100 in all tumor sub-regions. The difference between the FeTS segmentation result of the synthesized output before and after refinement, illustrated in Figure \ref{fig:refiner_result}, further proves the \textit{Refiner}'s notable performance in refining the tumor representation of the synthesized MR image. 

\subsubsection{Results on Hidden Test Set}
Our method was submitted through the Synapse platform, and the evaluation results on hidden test cases were reported as in Table~\ref{tab:final_result}. As we observed on our experimental results, SSIM score exhibited a slight decline; however, the Dice scores demonstrated significant performances, effectively capturing the 3D context of the tumor.
\begin{table}[t]
\caption{Dice score computed with respect to pseudo-label produced by FeTS model. It is computed for each tumor region, which includes the whole tumor (WT), enhancing tumor (ET), and tumor core (TC). IE (\textit{mean}) represents applying the intensity encoding method with sampling the target intensity as a mean value.}\label{tab1}
\centering
\begin{tabular}{l|c|c|c|c}
\hline
\multirow{2}{*}{Method}  & \multicolumn{4}{c}{Dice Score} \\ \cline{2-5}
                   & WT & ET & TC &Avg \\ \hline \hline
HF-GAN + IE (\textit{mean}) & 0.8133 & 0.6288 & 0.6841 & 0.7087 \\ \cline{1-5} 
+\textit{Refiner} & \textbf{0.9318} & \textbf{0.7358} & \textbf{0.7896} & \textbf{0.8190} \\ \hline
\end{tabular}
\label{tab:Dice}
\end{table}

\begin{table}[b]
\caption{Evaluation results on hidden test set. Dice Score and HD95 are computed separately for each tumor regions, while SSIM is computed for whole brain MR Image.}\label{tab1}
\centering
\begin{tabular}{l|c|c|c|c|c|c|c}
\hline
\multirow{2}{*}{Method}  & \multirow{2}{*}{SSIM}& \multicolumn{3}{c|}{Dice Score}  &  \multicolumn{3}{c}{HD95} \\ \cline{3-8}
                  & & WT & ET & TC &WT & ET & TC \\ \hline \hline
Mean & 0.8183 & 0.8399 & 0.7133 & 0.7572 & 17.4707 & 30.3729 & 28.1854 \\ \cline{1-8} 
Std & 0.0192 & 0.1949 & 0.2927 & 0.2985 & 47.2653 & 91.2351 & 81.1976 \\ \hline
Median & 0.8178 & 0.9076 & 0.8309 & 0.8984 & 7.5498& 3.0000 & 5.3852 \\ \hline
25th-quantile & 0.8051 & 0.8423 & 0.6271 & 0.7031 & 4.4721 & 1.0000 & 3.0000 \\ \hline
75th-quantile & 0.8308 & 0.9418 & 0.9269 & 0.9529 & 12.4593 & 8.8588 & 11.0226
\end{tabular}
\label{tab:final_result}
\end{table}

\section{Discussion}

Our proposed two-stage method has shown excellent MRI synthesis results while addressing essential problems of the image synthesis method. To reduce the complexity of 3D refinement, we first focus on synthesizing results with improved perceptual quality. As a result, the MRI synthesized from the 2D slices has consistent and accurate anatomical structures in most cases. However, as you can see in Figure~\ref{fig:refiner_result}, segmentation results of the synthesized MRI sometimes show poor results, even if the synthesized MRI is perceptually similar to the real MRI in the human eye. To address this problem, we propose the \textit{Refiner} for the second stage. Only subtle adjustments are applied to enhance the visibility of structural features for the segmentation model without significantly altering the overall structure, and this approach leads to substantial numerical improvements in terms of the Dice score, but minor decreases in terms of the SSIM score.

The decreases in SSIM score are likely attributed to the conflict between the segmentation metric and the SSIM metric. The SSIM metric evaluates how closely the synthesized output MRI resembles the ground truth MRI, while the Dice metric measures how similar the labels obtained from the output are to the ground truth segmentation labels. In essence, the Dice metric tends to favor outputs where tumor regions are more clearly identified as tumors and healthy regions are more distinctly recognized as healthy, even if its appearance deviates from the ground truth. For instance, intensity jitters in white matter area are regarded as unnecessary noise from the perspective of the Dice metric. This leads \textit{Refiner} to produce output with smoother white matter region by eliminating jitters, even though the jitters are present in the ground truth MRI. Thus, while the model trains to optimize the output's tumor representation, it resulted in output intensities that deviates from the ground truth, which explains the marginal decrease in SSIM scores.

We believe that the potential of the \textit{Refiner} extends beyond the current results. We plan to further evaluate its performance in tasks that require more significant modifications than those in the present study. By doing so, we aim to fully explore the capabilities of the \textit{Refiner} and its potential impact on more challenging applications. This future work will help better understand and demonstrate the \textit{Refiner}'s ability to handle more complex adjustments and its broader applicability in medical imaging.

\begin{credits}
\subsubsection{\ackname} This work was supported by Institute for Information \& communications Technology Promotion(IITP) grant funded by the Korea government (MSIT) (No.00223446, Development of object-oriented synthetic data generation and evaluation methods).
\end{credits}
%
%
%
\bibliographystyle{splncs04}
\bibliography{main}

\newpage
\appendix

\section{Implementation}
\subsection{First Stage}
We followed the baseline configuration~\cite{cho2024unified} with an improvement in the modality infuser using intensity encoding. Intensity priors are computed as a normal distribution (mean, standard deviation) as follows: (0.4332, 0.1003) for T1, (0.2498, 0.0626) for T2, (0.3034, 0.1017) for FLAIR and (0.2404, 0.0623) for T1ce. We optimized the network during 20 epochs with batch size 24, and the total loss $L_G$ is
\begin{equation}
    L_{G}=\alpha L_{rec}+\beta L_{sim}+\gamma L_{cyc}+\delta L_{adv}+\epsilon L_{cls}+\zeta L_{ssim}+\eta L_{ssim-tumor},
\end{equation}
where $L_{rec}$, $L_{sim}$, $L_{cyc}$, $L_{adv}$, $L_{cls}$, $L_{ssim}$ and $L_{ssim-tumor}$ represent the L1 reconstruction loss, cosine similarity loss, cycle-consistency loss, adversarial loss, classification loss, SSIM loss, and SSIM loss for tumor region, respectively, with $\alpha$, $\beta$, $\gamma$, $\delta$, $\epsilon$, $\zeta$ and $\eta$ set to 10, 1, 1, 0.25, 0.25, 5 and 5, respectively.

\subsection{Second Stage}
The number of channels in each layer of the encoder U-net, as well as the output channels and the channels for the query, key, and value vectors, were all set to 64. The total loss used for training \textit{Refiner} is as follows:
\begin{equation}
    L_{total} = \lambda_1 L_{ssim} + \lambda_2 L_{Dice},
\end{equation}
where $L_{ssim}$ represents SSIM loss and $L_{Dice}$ represents Dice loss of tumor label segmented from synthesized MR image using FeTS model.
The hyperparameters $\lambda_1$ and $\lambda_2$ were both set to 1.0 to ensure that the SSIM loss and the Dice loss had equal weight during training. The training was conducted for 100 epochs with batch size of 4. 

\end{document}